\documentclass[pra,twocolumn,longbibliography]{revtex4}
\usepackage{graphicx}
\usepackage{dcolumn}
\usepackage{bm}
\usepackage{physics}
\usepackage{amsmath}
\usepackage{array}
\usepackage{color}
\usepackage{times}

\newcolumntype{P}[1]{>{\centering\arraybackslash}p{#1}}
\usepackage[colorlinks=true, breaklinks=true, linkcolor=red, citecolor=blue, urlcolor=blue]{hyperref}

\newcommand{\bonnpi}{Physikalisches Institut, University of Bonn, Nussallee 12, 53115 Bonn, Germany}

\usepackage{tikz}
\usetikzlibrary{calc}
\usetikzlibrary{shapes}
\usepackage{ifthen}
\usepackage{booktabs}
\usepackage{multirow}

\begin{document}

\title{Level statistics of the one-dimensional dimerized Hubbard model}
\date{\today}

\begin{abstract}
    The statistical properties of level spacings provide valuable insights into
    the dynamical properties of a many-body quantum systems. We investigate the level statistics of the Fermi-Hubbard model with dimerized hopping amplitude and find that after taking into account translation, reflection, spin and $\eta$ pairing symmetries to isolate irreducible blocks of the Hamiltonian, the level spacings in the limit of large system sizes follow the distribution expected for hermitian random matrices from the Gaussian orthogonal ensemble.
We show this by analyzing the distribution of the ratios of consecutive level spacings in this system, its cumulative distribution and quantify the deviations of the distributions using their mean, standard deviation and skewness.
\end{abstract}
\author{Karin Haderlein}
\affiliation{\bonnpi}
\author{David J. Luitz}
\affiliation{\bonnpi}
\author{Corinna Kollath}
\affiliation{\bonnpi}
\author{Ameneh Sheikhan}
\affiliation{\bonnpi}

\maketitle

\section{Introduction \label{sec:intro}}

The dynamical behavior of quantum systems has attracted a lot of interest in the past years. Whereas the low temperature equilibrium properties of a system can be typically approximated by simplified models and only the low energy part of the model's spectrum is needed, the dynamics of a system is governed by arbitrary parts of the spectrum. Depending on the non-equilibrium situation, this can extend over the entire spectrum or only have contributions from parts of the spectrum. An important question after the perturbation of a system is typically whether and how fast it rethermalizes. The conditions under which isolated quantum many-body system thermalize or fail to thermalize have been actively investigated (see for example \cite{Srednicki1994,KollathAltman2007, PolkovnikovVengalattore2011,NandkishoreHuse2015,AbaninSerbyn2019,DalessioRigol2016} and references therein). It is well known that for example the presence of symmetries is responsible for the existence of conserved quantities \cite{BasdevantDalibardBook2002} and therefore influences drastically the dynamics of the system. It is the general opinion that only a generic chaotic system shows thermalization towards a suitable statistical mechanics ensemble~\cite{Haake2000}.

Many recent studies have pointed out examples which do not thermalize, such as many-body localization \cite{BaskoAltshuler2006,GornyiPolyakov2005,ZnidaricPrelovsek2008,PalHuse2010,LuitzAlet2015,NandkishoreHuse2015,AbaninSerbyn2019}, and Hilbert space fragmentation~\cite{SalaPollmann2020, KhemaniNandkishore2020, MoudgalyaRegnault2022}. The presence of a large number of conserved quantities or almost conserved quantities prohibits or slows down dramatically thermalization. More recently the attention has been attracted by the phenomenon of quantum many-body scarring \cite{ShiraishiMori2017,TurnerPapic2018,KhemaniChandran2019,HoLukin2019,MoudgalyaRegnault2022}  where  even in overall chaotic systems small subspaces are protected and fail to thermalize if the initial state has overlap with these subspaces~\cite{BernienLukin2017}.

The spectral properties of a model represent an important indicator whether it is expected to behave  in an integrable or chaotic manner.
In particular, the statistics of level spacings are used to characterize quantum many body models.  
Spectral gaps of one class of models follows the Poisson distribution and another class is described by random matrix theory statistics \cite{Mehtabook} as the two limiting cases. Non-trivial quantum integrable many body systems typically show Poisson statistics. In contrast, the random matrix theory statistics is conjectured to hold for generic chaotic systems \cite{BerryTabor1977, BohigasSchmit1984}. Characteristic for the corresponding spectra is the presence of energy level repulsion. For the random matrices different symmetries can be present. Which of the random matrix ensembles describes the features of the many body system well therefore also depends on the symmetries in the system.

Many different many body quantum systems have been investigated. To mention just a few, the prime examples are the Hubbard model~\cite{PoilblancMontambaux1993} and variants of it~\cite{JafariHosseinzadeh2020, DeMarcoKollath2022}, or the Bose-Hubbard model~\cite{KollathLaeuchli2010}, a kicked-parameter model of spinless fermions~\cite{Prosen1999}, or generalized Sachdev-Ye-Kitaev models~\cite{Haque2019}.

There are various quantum many body models which are central in condensed matter physics. Among these, the Fermi Hubbard model is well known and relatively simple, while capturing two major actions of fermions on a lattice. These consist of a particle hopping to a neighbouring site and an on-site interaction between two fermions of opposite spin. Several aspects of the Fermi Hubbard model have been well studied  \cite{Gebhard1997}. In particular its one-dimensional version~\cite{EsslerKorepin2005, Giamarchibook} has been pointed out to be Bethe-Ansatz integrable and its level statistics have been found to follow the Poisson statistics \cite{PoilblancMontambaux1993}, in agreement with the conjecture mentioned before. In contrast, the two-dimensional Hubbard model at low filling exhibits the statistics of the Gaussian orthogonal ensemble (GOE)~\cite{Bruus1996} from the random matrix theory.

One possible generalization of the Hubbard model is to add alternating hopping amplitudes between two sites, thus doubling the unit cell of the lattice. This so-called dimerized Hubbard model has attracted a lot of attention during the past \cite{PencMila1994b, LeGinossar2020, WangWu2015, YeFan2016, BarbieroGoldman2018} and more recently the interest has been revived due to its topological properties. In particular, the non-interacting model corresponds to the SSH model \cite{SuHeeger1979} which has been intensively studied in recent years \cite{AsbothPalyi2016, XieYan2019}. 

In this work, we investigate the spectral properties of the dimerized Hubbard model. The presence of the so-called $\eta$ symmetry \cite{Esslerbook,KitamuraAoki2016}  makes this model special. The symmetry leads to a tower of equally spaced many body eigenstates due to the existence of a spectrum generating algebra \cite{MoudgalyaBernevig2020}. We find that only after taking into account all important symmetries including this $\eta$-symmetry, the level statistics within irreducible symmetry sectors are clearly of GOE nature in a wide range of considered parameters. 

In Sec.~\ref{sec:model} we introduce the dimerized Hubbard model studied in this paper. In Sec.~\ref{sec:LS} the level statistics of the spectrum of many body systems and their properties are presented. In order to calculate the level statistics correctly, one must consider the important symmetries present in the model, which are explained in Sec.~\ref{sec:sym}. The energy spectrum is computed using exact diagonalization, taking into account some of the symmetries explicitly. In Sec.~\ref{sec:methods} we explain how to separate the symmetry sectors of the remaining symmetries  for the level statistical analysis. The spectral analysis of the model for different parameters and different fillings is presented in Sec.~\ref{sec:results}.

\section{Model \label{sec:model}}

In this work we study the dimerized Fermi Hubbard model. In contrast to the standard Hubbard model it contains alternating hopping amplitudes $t$ and $t'$ changing every other site. The corresponding Hamiltonian of the dimerized Hubbard model is given by
\begin{align}
\label{eq:DimerHubb}
H =& -t\sum_{\sigma}\sum_{\substack{j = 1 \\ \mathrm{odd}}}^{L-1} \left(c^\dagger_{j, \sigma} c^{\phantom{\dagger}}_{j+1, \sigma}+c^\dagger_{j+1, \sigma} c^{\phantom{\dagger}}_{j, \sigma}\right) \nonumber\\
&-t^\prime\sum_{\sigma}\sum_{\substack{j = 2 \\ \mathrm{even}}}^L \left(c^\dagger_{j, \sigma} c^{\phantom{\dagger}}_{j+1, \sigma}+c^\dagger_{j+1, \sigma} c^{\phantom{\dagger}}_{j, \sigma}\right) \nonumber\\
&+ U \sum_{j = 1}^{L}n_{j, \uparrow}n_{j, \downarrow},	
\end{align}
where $c^{\phantom{\dagger}}_{j, \sigma}$ is the fermionic annihilation operator and $n_{j, \sigma} = c^\dagger_{j, \sigma}c^{\phantom{\dagger}}_{j, \sigma}$ the particle number operator at site $j$ with spin $\sigma$. We sum over $\sigma \in \{\uparrow, \downarrow\}$. We use periodic boundary conditions identifying $c^{\phantom{\dagger}}_{L+1 ,\sigma}$ with $c^{\phantom{\dagger}}_{1, \sigma}$ where $L$ is the length of the chain and is an even number.
The first two lines in the Hamiltonian correspond to the kinetic contributions with the alternating hopping amplitudes $t$ and $t'$. It is the Su-Schrieffer-Heeger (SSH) model which is known for its topologically non-trivial phase \cite{SuHeeger1979, AsbothPalyi2016}.

Due to the alternating hopping the unit cell of the lattice contains two sites and the dimerization opens a band gap which leads to an insulating behaviour at quarter filling. 

The last term  in Eq.~\eqref{eq:DimerHubb} contains the on-site interaction between two particles of opposite spin with interaction strength $U$ which we typically choose repulsive, i.e.~$U>0$. 
The ground state phase diagram of the dimerized Hubbard model is well studied \cite{PencMila1994b,TsuchiizuSuzumura2001, BenthienJeckelmann2005} and also the edge states have been investigated using bosonization \cite{JinGiamarchi2023}
.

It comprises at zero dimerization the phases of the standard Hubbard model reaching from a metallic state at incommensurate filling to an antiferromagnetic Mott insulator at half filling. Adding the dimerization leads again to an additional insulating phase at quarter filling.

\section{Level statistics}
\label{sec:LS}
In this section we introduce the properties of the spectral statistics of many body systems. The level statistics of many body quantum systems are often used to identify the integrability of the model. This is based on the conjecture that the spectral statistics of quantum many body systems exhibit different statistics \cite{BerryTabor1977, Mehtabook}; if a system is chaotic, it is believed to exhibit universal features of random matrix theory, whereas if a system is integrable, its level statistics follow the Poisson distribution. There are many examples where spectral analysis has been performed and has proven to be very useful in identifying characteristics of these many-body systems \cite{PoilblancMontambaux1993, KollathLaeuchli2010, VyasSeligman2018, DeMarcoKollath2022}.

The level spacing of a many body spectrum is defined as the distance between adjacent many body eigenvalues
$$\delta_j = E_{j+1}-E_j,$$ where the eigenenergies $E_j$ are sorted in ascending order. The distribution of the level spacings $\delta_i$ is conjectured to follow different distributions for the two cases, namely integrability or chaos. In the case of an integrable model the level spacing should follow a Poisson distribution \cite{BerryTabor1977}
\begin{align}
\label{eq:PdeltaPoisson}
P_{\mathrm{Poisson}}(\delta/\Delta)&= \exp(-\frac{\delta}{\Delta})
\end{align}
where $\Delta$ is the mean value of the level spacings. In contrast, under the condition that the energy spectrum is divided into all existing simultaneously commuting discrete symmetry sectors, the distribution of the level spacings in each symmetry sector follows the behavior of a random matrix ensemble. In the case of the dimerized Hubbard model, the system has time-reversal and rotational symmetries. Such a system follows the distribution obtained for the Gaussian Orthogonal Ensemble (GOE) given approximately by
\begin{align}
\label{eq:PdeltaGOE}
P_{\mathrm{GOE}}(\delta/\Delta) &= \frac{\pi}{2}\frac{\delta}{\Delta} \exp(-\frac{\pi}{4}\frac{\delta^2}{\Delta^2}) \nonumber.
\end{align}
Here the distribution is calculated using approximations in the random matrix theory \cite{Mehtabook}.

However, an analysis of the level spacings requires an unfolding procedure to determine $\Delta$ which is not trivial to perform. Therefore, it is more convenient to use the ratios \cite{OganesyanHuse2007} of consecutive level spacings

\begin{equation}
    r_j = \frac{\min(\delta_j, \delta_{j+1})}{\max(\delta_j, \delta_{j+1})}. 
\end{equation}

The energy spectrum is highly degenerate because there are many symmetries in this model. 
When there are $n$ degenerate energies in the spectrum, we have $n-1$ zero-value level spacings ($\delta_j=0$) in a row. We don't consider these degenerate energies in the calculation of distribution function. 

\begin{figure}
	\centering
	\includegraphics[width = 0.49\textwidth]{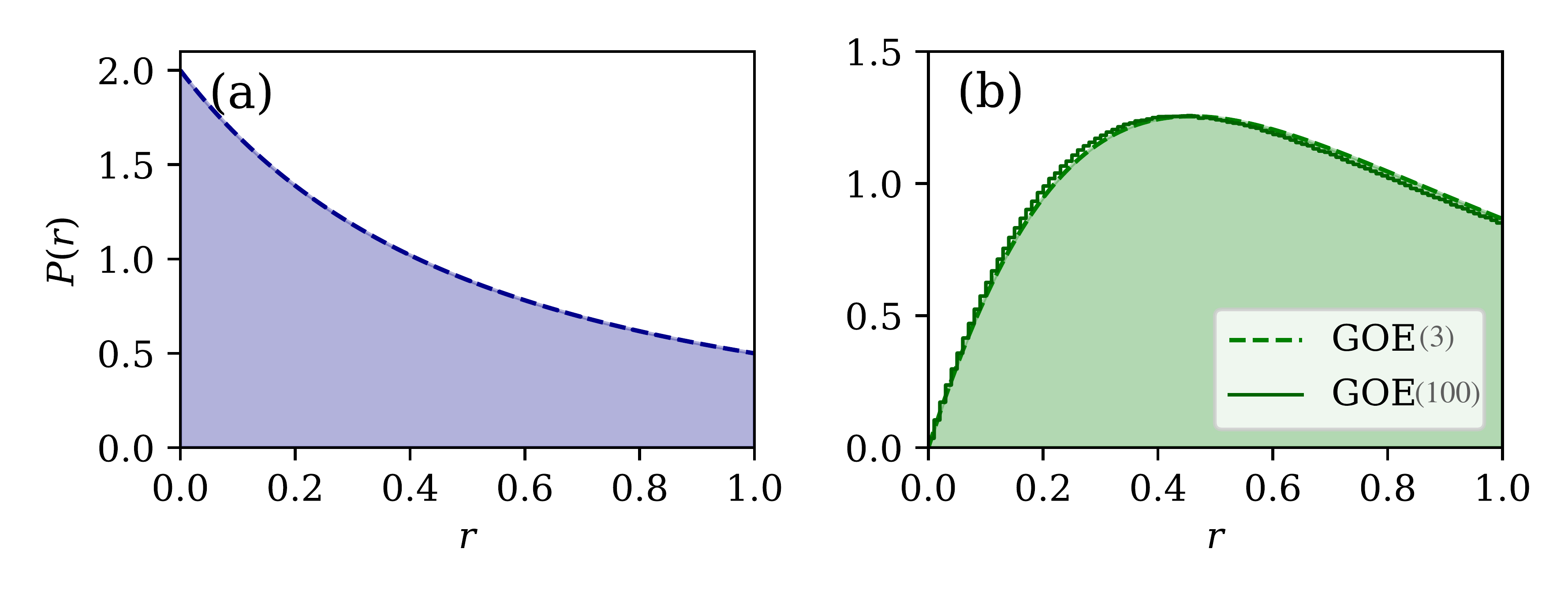}
	\caption{Predictions for the distribution of ratios $P(r)$ for (a) the Poisson statistics and (b) the GOE statistics. For the GOE statistics both the analytic prediction for $D = 3$ matrices and the numerical results for $D = 100$ matrices are displayed. The deviations are minor such that we use the numerical results. }
	\label{fig:theor}
\end{figure}

The distribution of the consecutive level spacings for random $3\times3$ GOE matrices can be analytically calculated and is given by \cite{AtasRoux2013}
\begin{align}\label{eq:PoisGOE}
    P_{\mathrm{GOE(3)}}(r) &= \frac{27}{4}\frac{r+r^2}{(1+r+r^2)^{5/2}}, \nonumber\\
P_{\mathrm{Poisson}}(r)& = \frac{2}{(1+r)^2}.
\end{align} 
These functions are depicted in Fig.~\ref{fig:theor}.

The mean value of the level spacings ratios $r_j$ for these  analytically calculated distributions (Eq.~\eqref{eq:PoisGOE})
is given by \cite{AtasRoux2013}
\begin{equation*}
    \begin{split}
        \expval{r}_{\mathrm{GOE(3)}} &= 4-2\sqrt{3} = 0.535898384862245\dots \quad \text{and} \\ 
        \expval{r}_{\mathrm{Poisson}} &=2\ln 2 -1 = 0.386294361119890\dots
\end{split}
\end{equation*}

 Let us note, that here we give a high precision for reference, which in principle we do not need for the comparison with our numerical result. For larger GOE matrices of size $N\times N$, there are no analytical results available, but the distribution and its moments can be estimated numerically. The distribution converges rapidly with $N$ to a universal result for infinite random matrices and we find that $N=100$ is already representative of this limit.

We therefore show the numerical results calculated for random GOE matrices of dimension $N = 100$ to Fig. \ref{fig:theor}. 
The difference of the GOE(3) and the GOE(100) ensembles are small but visible and we therefore use the universal GOE(100) ensemble for more accurate comparisons. The mean value calculated for the GOE(100) \footnote{These estimates were calculated by averaging over $7.8\cdot10^6$ random matrices using the central 80\% of the spectrum} is $\langle r \rangle_{\mathrm{GOE}(100)} = 0.530695\pm0.000012$ which is slightly lower than the result for GOE(3) $\langle r \rangle_{\mathrm{GOE(3)}}$, a trend visible in the distribution in Fig. \ref{fig:theor}.

Further descriptive statistics of the distribution of the gap ratios $r$ of the GOE($100$) ensemble are the standard deviation $\mathrm{std}_{\mathrm{GOE}(100)}(r)=0.25444(1)$ and the skewness $\mathrm{skew}_{\mathrm{GOE}(100)}(r)=0.0203(1)$.

The numerically determined distribution of $P(r)$ requires the choice of bins and the recording of histograms.
In order to exclude any potential bias from binning, we also consider the cumulative density function $\mathrm{CDF}(r)$ given by the probability to observe gap ratios less then $r$: 
\begin{equation}
    \mathrm{CDF}(r) = \int \limits_0^r \mathrm{d}x \, P(x).
    \label{eq:CDF}
\end{equation}
The main advantage of this quantity is that it can be represented 
without any binning by approximating $\mathrm{CDF}(r)$ by a step function with steps of size $1/N$ at each of the observed $N$ data points $r_j$. For large amounts of data, this function becomes very smooth and we can consider the difference of $\mathrm{CDF}(r)$ with its estimate for the random matrix ensemble $\mathrm{GOE}(100)$ to quantify the convergence of the entire distribution.
For the Poisson distribution, this function is given by
\begin{equation}
    \mathrm{CDF}_\mathrm{Poisson}(r) = 2 - \frac{2}{r+1}.
    \label{eq:CDF_Poisson}
\end{equation}

\section{Symmetries \label{sec:sym}}
The dimerized Hubbard model possesses several symmetries. We discuss in the following the ones which are important for the spectral analysis. 

\paragraph{Translation Symmetry:}
Due to the bipartite lattice, a translation of the local creation and annihilation operators by two sites with operator $T_2$ as 
\begin{equation}
T^{\phantom{\dagger}}_2c_{j, \sigma}T_2^\dagger = c_{j+2, \sigma}
\end{equation}
leaves the Hamiltonian invariant. Translationally invariant states can be identified by their momentum $k = \frac{4\pi l}{L}$ with $l = 0, \dots, L/2 -1$. We describe in appendix \ref{app:translation} in more detail how to implement the translational symmetry. 

\paragraph{Reflection Symmetry:}
The dimerized Hubbard model is invariant under reflection $R$ of the chain around the center of a bond, i.e.~for the central bond this is given by
\begin{equation}
R c_{j, \sigma} R^\dagger = c_{L+1-j, \sigma}.
\end{equation}
As applying the reflection operator twice is an identity operator ($R^2=I$) the corresponding states are described by their parity $p =\pm 1$.
Note, that in general translation and reflection symmetry do not commute. For the reflection around the central bond, it is only for $k = 0, \pi$ possible to find states both invariant under translation and reflection. 

\paragraph{Spin rotation symmetry:}
The dimerized Hubbard model additionally possesses a spin rotation SU(2) symmetry, i.e.~ the direction of the spin can be rotated to any direction. The corresponding operators for the symmetry are the spin operators defined by
\begin{equation}
S_\alpha = \frac{1}{2}\sum_{j = 1}^L \sum_{a,b \in \uparrow,\downarrow}c_{j,a}^\dagger (\sigma_\alpha)^{\phantom{\dagger}}_{a,b}c^{\phantom{\dagger}}_{j,b}
\end{equation}
where $\sigma_\alpha$ denote the Pauli matrices with $\alpha=x,y,z$ (here $\hbar=1$). These spin operators commute with the Hamiltonian. Since the spin operators do not commute with themselves, we choose the square of the total spin $S^2=S_x^2+S_y^2+S_z^2$ and its $S_z$ component with the respective eigenvalues of $s$ and $m$. The eigenvalues $s$ and $m$ correspond to conserved quantum numbers. 
Note, that besides the $S_\alpha$'s and $S^2$ operators, also the ladder operators $S_\pm=\frac{1}{2}(S_x\pm S_y)$ commute with the Hamiltonian
\begin{equation}
	\comm{H}{S_\pm} = 0.
\end{equation}
These ladder operators will be used in order to associate the many body energy eigenvalues to the values of the total spin $s$ as described later.

\paragraph{$\eta$-pair symmetry:}
The Hubbard model and also the dimerized Hubbard model has additionally the so-called $\eta$-pair symmetry \cite{EsslerKorepin2005, MoudgalyaBernevig2020}. This is an additional SU(2) symmetry generated by the $\eta$-pair operators
\begin{align}\label{eq:etadef}
&\eta_+ = \sum_{j = 1}^L(-1)^{j + 1}c^\dagger_{j, \uparrow}c^\dagger_{j, \downarrow}, \quad \eta_- = \eta_+^\dagger = \sum_{j = 1}^L(-1)^{j + 1}c_{j, \downarrow}c_{j, \uparrow} \nonumber\\
&\text{and} \quad \eta_z = \frac{1}{2}(\hat{N}-L)
\end{align}
where $\hat{N}$ is the total particle number operator. The ladder operators can be interpreted as creation or annihilation of $\eta$-pair states. Due to its similarity to the spin symmetry it is also referred to as the pseudospin.
Here, the two independent operators defining the independent conserved quantum numbers are given by $\eta^2$ with conserved quantity $\xi$ and $\eta_z$ with $\tilde{n}$. From Eq.~\eqref{eq:etadef} one can see that the quantum number associated to the $z$-component of the $\eta$ operators is related to the total particle number $N$ and system length $\tilde{n}=\frac{1}{2}(N-L)$.
Let us note, that the commutation relations of the ladder operators with the Hamiltonian are given by
\begin{equation}\label{eq:comUeta}
	\comm{H}{\eta_\pm} = \pm U \eta_\pm.
\end{equation}
Unlike to the spin ladder operators, in the presence of an interaction $U\neq 0$, the commutator of the $\eta$-ladder operators does not vanish, but is a finite value times the operators themselves. This property is often called a spectrum generating algebra \cite{MoudgalyaBernevig2020}. This property is important for the spectrum of the system, since it causes a series of equally spaced many body energy eigenstates with energy distance $U$. If $\Psi_i$ is an eigenstate of $H$ with energy $E_i$, the $\eta$-pair creation operator $\eta_+$ generates another eigenstate $\eta_+ \Psi_i$ with eigenenergy $E_i+U$. Thus,  by iterating this procedure an entire tower of equally spaced eigenvalues can be generated using the creation operator. In the next section we describe how to remove these eigenstate towers before analysing the properties of the spectrum.

\section{Methods \label{sec:methods}}
For level statistics analysis it is necessary to find a spectrum divided into its symmetry subsectors. If symmetries are not taken into account, the distribution may approach the Poisson distribution and exhibit additional features not present in the GOE-like distribution  \cite{RosenzweigPorter1960,BerryRobnik1984,SunLiu2020,GiraudAlet2020}. This was investigated for independent symmetries by Giraud and co-workers \cite{GiraudAlet2020}. They presented analytical results for the distributions of random matrices comprised of several independent symmetry blocks, corresponding to the presence of a discrete symmetry\cite{GiraudAlet2020}.

\begin{figure}
	\centering
	\includegraphics[width = 0.49\textwidth]{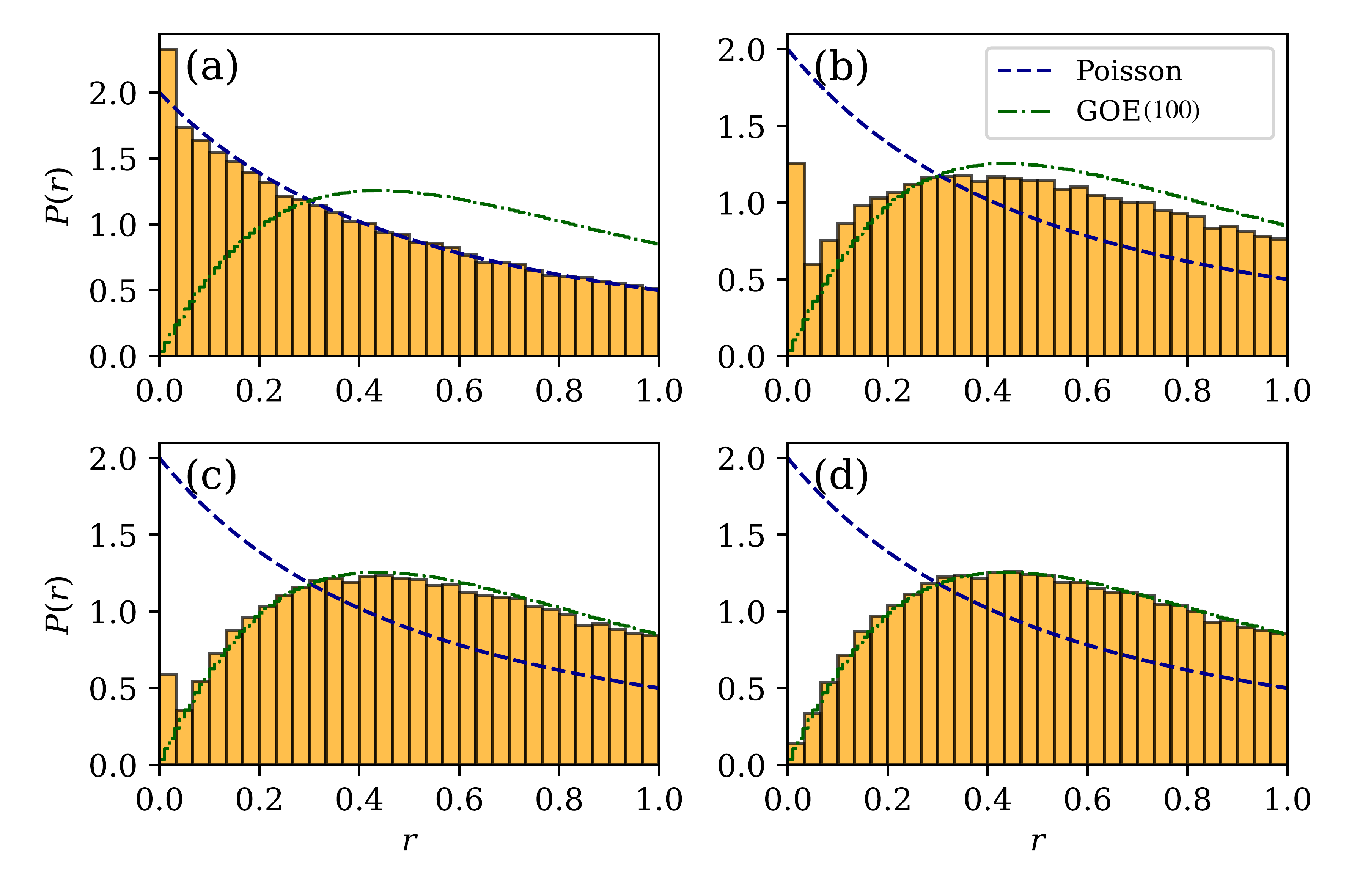}
	\caption{Distribution of ratios $P(r)$ for $L = 16$, $N = 8$ and $m = 3$ for $t'/t = 2$, $U/t = 2$. In the analysis, the spectrum was divided into different symmetry subsectors considering only certain symmetries in order to visualize the importance of these symmetries. For all subpanels the quantum number $m$ and $N$ where fixed. For (b) additionally the blocks where separated considering the momentum $k$. For (c) the momentum $k$, parity $p$ and total spin $s$ were taken into account and  for (c) the momentum $k$, parity $p$, total spin $s$ and total pseudospin $\xi$. The distribution in (a) resembles a Poisson like distribution. By dividing the spectrum into more symmetry sectors as done in (b) and (c) the distribution approaches more the GOE prediction. In (d) the spectrum was divided into all known symmetry subsectors and the distribution matches the GOE distribution well.  }
	\label{fig:symsec}
	\end{figure}

For our model, we illustrate the consequences of neglecting symmetries in Fig.~\ref{fig:symsec} by also showing the distribution for the case where not all symmetries are considered. In the previous section we found that for the considered dimerized Hubbard model an independent set of good quantum numbers is given by the momentum $k$, parity $p$ (for $k=0,\pi$), total spin $s$, spin $z$-component $m$, total pseudospin $\xi$ and pseudospin $z$-component $\tilde{n}$. The employed exact diagonalization code implements explicitely the translation symmetry, reflection symmetry --where applicable--, the magnetization along the $z$-direction, $m$, and the total particle number $N$ or equivalently the pseudospin $z$-component $\tilde{n}$. However, in order to show the effect of the symmetries, we display in Fig.~\ref{fig:symsec} (a) the distribution of ratios for subsectors only divided into the $N$ and $m$ sectors. In this case for large values of $r$, the distribution resembles closely a Poisson distribution with an additional peak at small $r$. Beside the peak at small $r$, this would incorrectly lead to the conclusion that the system follows an integrable statistic.
By dividing the spectrum also into the translational symmetry blocks Fig.~\ref{fig:symsec} (b) the distribution drastically changes and becomes closer to the GOE like distribution. However, still considerable deviations can be seen and the  additional low $r$ peak is evident. 
The division into the subsectors of the total spin $s$ and the total pseudospin $\xi$  is performed by a postprocessing of the obtained data as described below.
In Fig~\ref{fig:symsec} (c) one sees that the consideration of the spin symmetry brings the distribution even closer to the GOE distribution. The largest deviation is still seen at low $r$. This deviation is caused by the $\eta$ pair symmetry and vanishes if also the towers of eigenstates are removed by postprocessing the data as seen in Fig~\ref{fig:symsec} (d). The resulting distribution is closely resembling the distribution expected for the GOE statistics.

In the following we describe the postprocessing used to separate the $s$ sectors and to remove the towers created by the $\eta$-symmetry. 
The postprocess for the spin $s$ symmetry is achieved by utilizing the commutation relations between the corresponding ladder operators and the Hamiltonian. Let us consider a state $\ket{s, m}$ with energy $E_s$ where  $m \in \{-s, -s+1, \dots, s-1, s\}$. Due to the commutation relation $\comm{H}{S_\pm} = 0$ a state $\ket{s, m'}$ has the same energy $E_s$ causing a (multiple) degeneracy. 
For the postprocessing procedure we begin with the maximum spin $z$-component $m=m_{\textrm{max}}$ sector. The energies in this sector belong to the spin $s = m_\textrm{max}$ sector. Now consider the $m_{\textrm{max}}-1$ sector. It contains energies from the $s = m_{\textrm{max}}$ and $s = m_{\textrm{max}}-1$ sectors. However, we already know the energies of the $s = m_{\textrm{max}}$ sector from the previous step. Hence, by omitting these degenerate energies we obtain the energies belonging to the $s =m_{\textrm{max}}-1$ subsector. This postprocess has to be continued until the desired total spin value $s$ is reached.

\begin{figure}
	\centering
	\includegraphics[width = 0.4\textwidth]{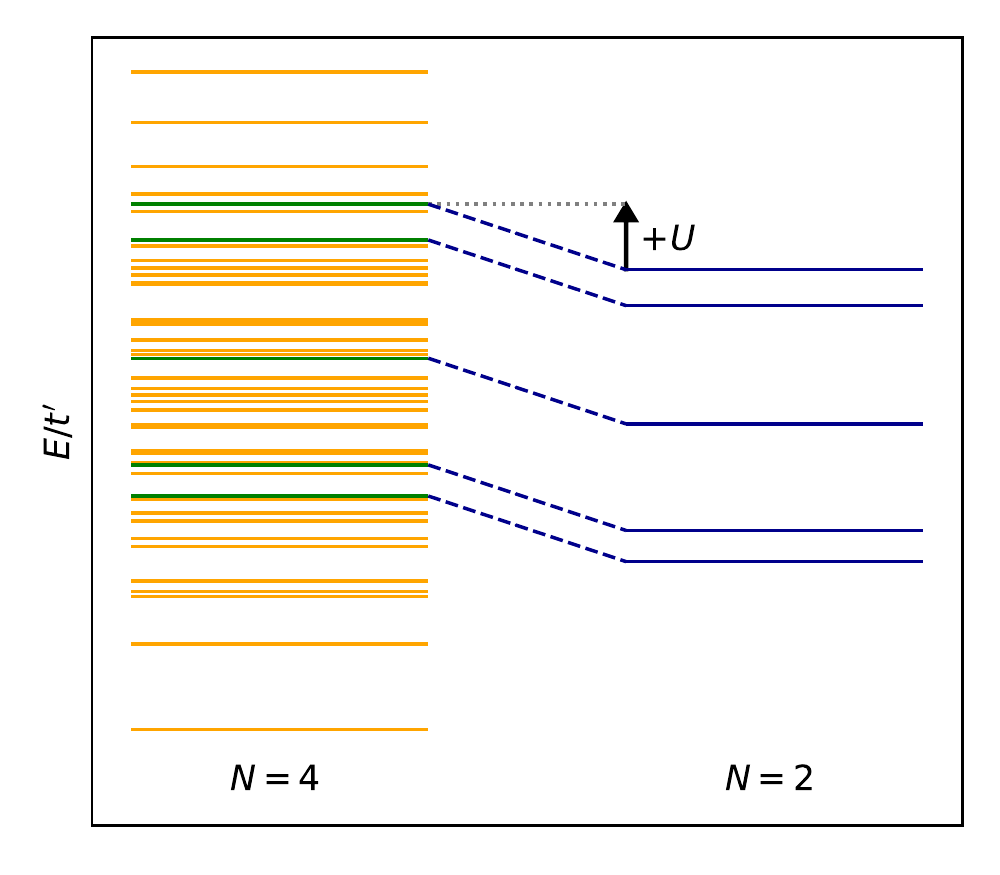}
	\caption{Energy spectra for $L = 4$, $m = 0$, $k = 2$ with $N = 4$ ($\tilde{n}=0$), $p = -1$ on the left hand side and $N = 2$ ($\tilde{n}=-1$), $p = 1$ on the right hand side. Moreover, $t'/t = 1.8$ and $U/t = 2$. The shifted energies by $U$ in the $N = 2$ sector coincide with energy levels of the $N = 4$ levels (green lines).}
	\label{fig:Ushift}
\end{figure}

For the pseudospin the post processing procedure is similar. However, the ladder operator does not commute with the Hamiltonian, since it forms a spectrum generating algebra, (Eq.~\eqref{eq:comUeta} ). As a consequence, two states $\ket{\xi, \tilde{n}}$ and $\ket{\xi, \tilde{n}'}$ do not have degenerate energies but energies shifted by the value $U$. This has to be considered in the postprocessing. Again, we start with the maximum pseudospin $z$-component $\tilde{n} = \tilde{n}_\mathrm{max}=1/2(N-L)$ which belongs to the sector with pseudospin $\xi = \tilde{n}_\mathrm{max}$.  We now consider the $\tilde{n}_\mathrm{max}-1$ sector. Due to the different commutation relation we find that the energies in the $\tilde{n}_\mathrm{max}$ sector are shifted by $U$ in the $\tilde{n}_\mathrm{max}-1$ sector. The remaining energies then form the symmetry sector with  pseudospin $\xi = \tilde{n}_\mathrm{max}-1$, which can be repeated until the sector with the desired pseudospin $\xi $ is reached. This is depicted in Fig.~ \ref{fig:Ushift}.

\section{Spectral analysis}
\label{sec:results}

In this section we analyse the spectra obtained for different parameter sets varying in particular the ratio of the hopping amplitudes and the interaction strength. Within the sectors of the symmetries we determine the ratios $r_j$ and consider their distribution. Our main finding of this paper is that for typical parameters the level statistics of the dimerized Hubbard model follow the statistics found for the GOE ensembles. We consider different system sizes up to $L=16$ and analyze the finite size effects. In order to decrease statistical fluctuations, the ratios $r_j$ are typically computed in all trackable subsectors separately and then averaged over different subsectors. This usually includes the combination of all $k$ and $p$ sectors. 

\subsection{Dependence on the dimerization of the hopping amplitude}
\begin{figure}[h]
	\centering
		\includegraphics[width = 0.49\textwidth]{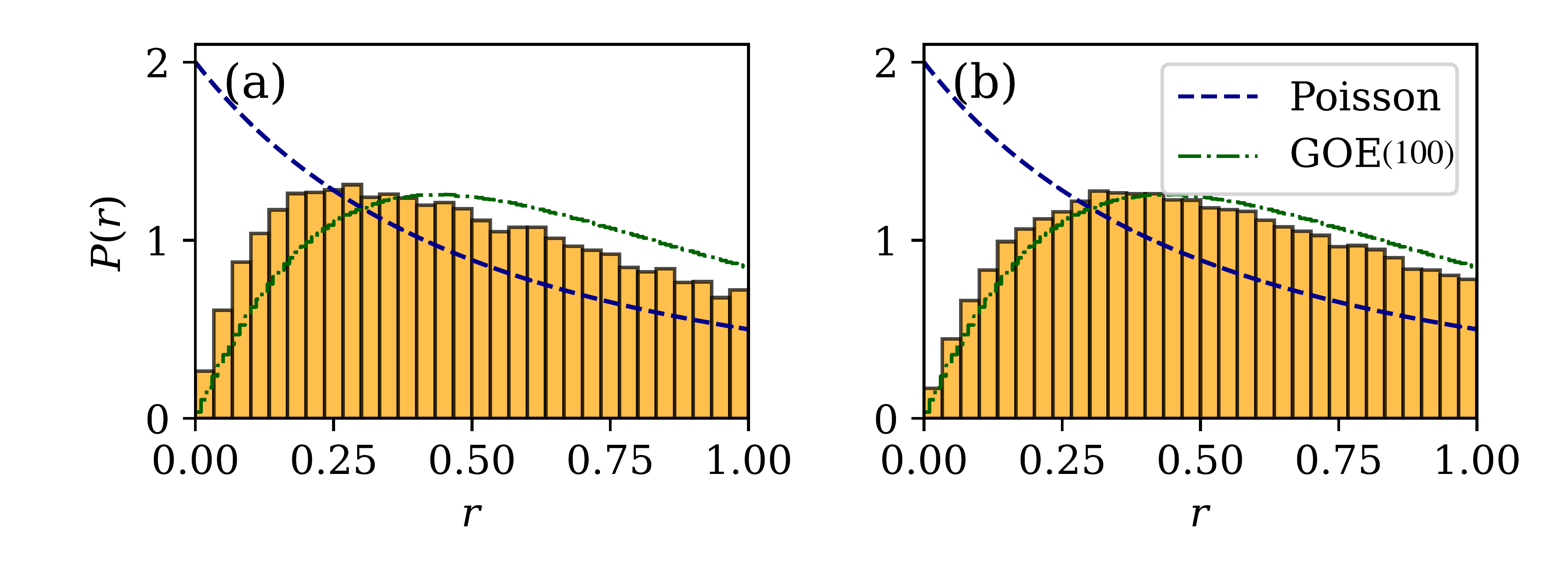}
	
	\caption{Distribution of ratios $P(r)$ for (a) $L = 12$, $N = 6$, $\xi = 3$ and (b) $L = 16$, $N = 8$, $\xi = 4$. Histograms are made for $t'/t = 1.1$ and $U/t = 2$. Moreover, all $k$ and $p$ sectors are merged after calculating the ratios. Additionally, for the smaller size $L = 12$ all available (interacting) spin $s= m$ sectors are used.
        }
	\label{fig:stagHubb_qf_L_11_comp}
\end{figure}

We start by showing a typical distribution at  $t'/t = 1.1$ at $U/t=2$ in Fig.~\ref{fig:stagHubb_qf_L_11_comp} for two different system sizes in comparison to the GOE(100) ensemble. The distributions fit relatively well to the GOE predictions and are clearly distinct from the Poisson distribution. For the considered system sizes, finite size effects still exist. The distribution for the smaller system size $L=12$ has its maximum much closer to $0$ than the larger system size. Thus, the remaining deviations from the GOE distribution decrease with increasing system size. We expect the GOE distribution to be reached for infinite system size.  

\begin{figure}[h]
	\centering
	
	\includegraphics[width =0.49\textwidth]{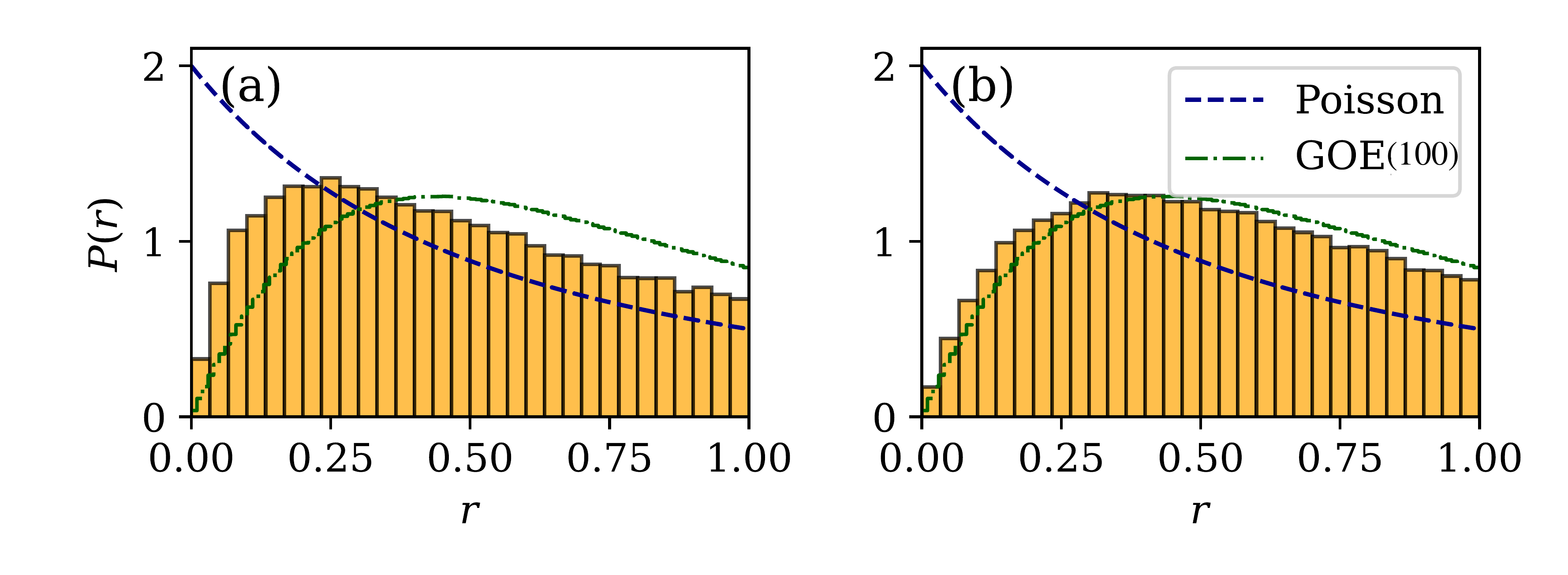}
	\caption{Distribution of ratios $P(r)$ for (a) $t'/t = 1.05$ and (b) $t'/t = 1.1$. Here the parameters are $L = 16$, $N = 8$, $s = m = 3$, $\xi = 4$ with $U/t = 2$. The ratios are calculated in $k, p$ sectors and all values of $k$, $p$ are considered. }
	\label{fig:stagHubb_L16N8_tp_comp}
\end{figure}

A similar shift of the distribution is found when decreasing the ratio of the hopping amplitudes to $t'/t = 1.05$ (Fig.~\ref{fig:stagHubb_L16N8_tp_comp}). Here again the maximum of the distribution shifts to lower values  for finite sizes. In this case where the deviation from the GOE distribution are stronger, we find that the finite size effects for this case are even more pronounced shifting the distribution towards the GOE distribution for larger system sizes. 

\begin{figure}[h]
	\centering
	
    \includegraphics[width = \columnwidth]{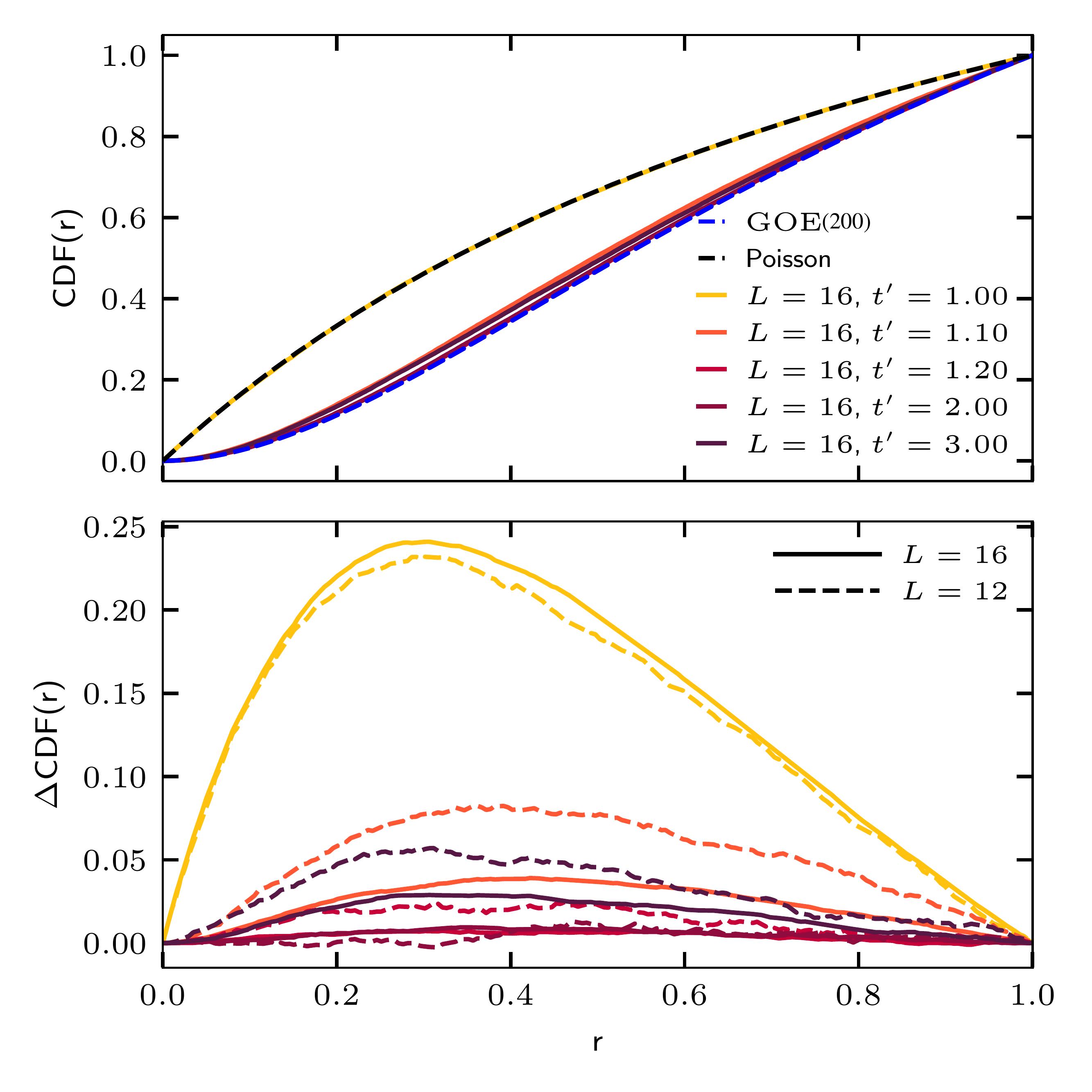}
    \caption{Cumulative distribution (CDF) of the ratios $r$ for different values of $t'/t$. The top panel shows the CDF for the parameters are $L = 16$, $N = 8$, $s = m = 3$, $\xi = 4$ with $U/t = 2$. The ratios are calculated in $k, p$ sectors and all values of $k$, $p$ are considered. The bottom panel displays the difference of the CDF with respect to the result for the GOE($200$) ensemble. Dashed lines show additional data for $L=12$, with $N=6$ and $s=m=0$ to illustrate the trend with growing system size. }
	\label{fig:cdf}
\end{figure}

In order to investigate the dependence on the ratio of the hopping amplitudes in more detail, we consider first the CDF defined in Eq. \eqref{eq:CDF} of the ratios $r_j$ and its difference to the CDF of the GOE. As seen in Fig.\ref{fig:cdf} (a), the cumulative distribution converge very quickly towards the CDF of the GOE and only small deviations are visible, while the integrable case of $t'=1$ shows clean Poisson statistics. Focusing on these deviations in  Fig.\ref{fig:cdf} (b), we find that even though these are still relatively pronounced at small and large  value of $t'/t$, they decrease considerably with increasing system size (cf.~$L=16$). This supports our previous findin that the expected behaviour of the GOE ensemble will arise in the thermodynamic limit.

\begin{figure}[h]
	\centering
	\includegraphics[width = 0.48\textwidth]{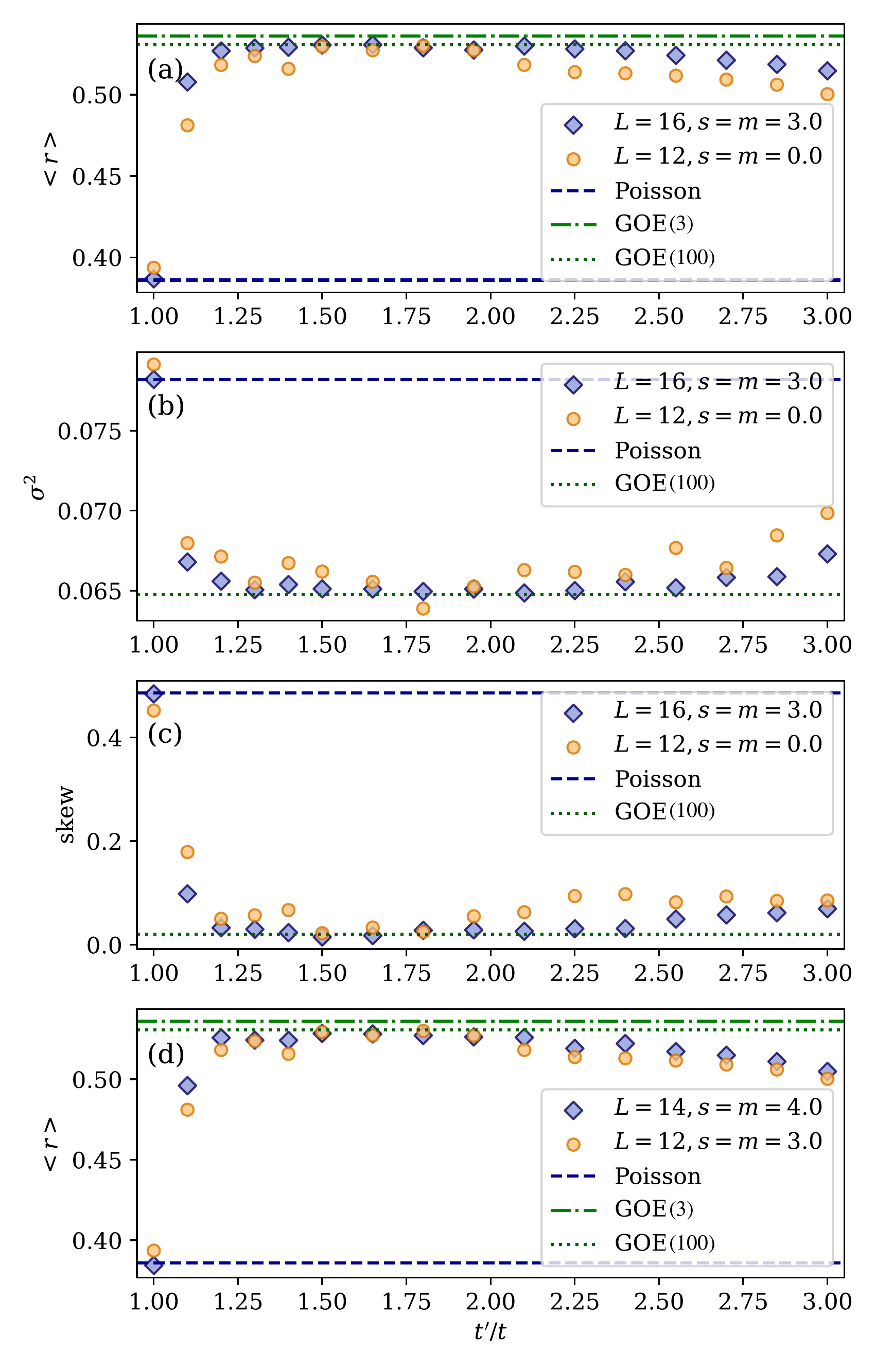}
	\caption{Mean values $\langle r\rangle$ of the distribution of ratios $P(r)$ for (a) quarter and (d) half filling as well as variance $\sigma^2$ (b) and skewness (c) depending on the hopping amplitude $t'/t$ with $U/t = 2$. For both system sizes the $k$, $p$ sector statistics are used and we fix $\xi = 3$ and $\xi = 4$ for $L = 12$ and $L = 16$, respectively. For the expected mean value of the GOE ensemble, we use the analytical predictions considering a dimension of 3 (green dot-dashed line) and a numerical prediction for a dimension of $D=100$ (green dotted line).  The results approach the GOE mean ratio over a wide range of hopping $t'/t$.}
	\label{fig:stagHubb_mr_tp}
\end{figure}

To quantify this further we also consider the mean values $\langle r\rangle=\sum_jr_j/N$, the variance $\sigma^2=\langle r^2\rangle-\langle r\rangle^2$, and the skewness $\langle \left(\frac{r - \langle r\rangle}{\sigma} \right)^3\rangle $ of the distribution of ratios $P(r)$ versus different hopping amplitudes $t'/t$ in figure \ref{fig:stagHubb_mr_tp}. Here, $N$ is the number of $r_j$ values.  We choose for the interaction strength $U/t = 2$. For quarter filling, we display the two system sizes $L = 12$ and $L = 16$ and the mean value the variance and the skewness in order to characterize the distributions. Since we find a similar behaviour for all the three moments of the distribution statistics, we only present the mean value for the case of half filling and $L = 12$ and $L = 14$. Note, that the spin is different for both system sizes as this was the lowest achievable spin by the exact diagonalization calculations. For $t'/t = 1$ the model coincides with the standard integrable Fermi Hubbard model. Therefore, we expect to find Poisson behaviour. We indeed find for this case that the mean ratio $\langle r\rangle$ agrees very well with the expected value for the Poisson distribution (blue dashed line). With increasing dimerization of the hopping amplitudes the mean ratio approaches the predictions of the GOE case (green dotted line).  
 However, for larger hopping amplitudes we find that the mean value start deviating again from the prediction of the GOE ensemble. In this limit, the strong dimerization favours the formation of almost decoupled dimers giving rise to the observed behaviour. 
 A similar behaviour of the approach of the expected value for the GOE is found for the variance and the skewness. That also this higher moments of the distribution show a similar behaviour supports that the entire distributions become close to the GOE distributions in the intermediate parameter regime.

 To verify whether the deviations from the GOE distribution are caused by finite size effects we show two different system sizes in figure \ref{fig:stagHubb_mr_tp}. We observe that the larger system size lies typically closer to the GOE prediction over a wider range of dimerizations. We may observe this exemplarily for the hopping amplitude $t'/t = 1.1$ for quarter filling in figure \ref{fig:stagHubb_qf_L_11_comp}. Comparing the distributions of ratios for the two system sizes $L = 12$ and $L = 16$ explicitly we observe that the larger system size resembles the GOE predictions more. Our results are in good agreement with the expectation that in the thermodynamic limit the distributions would approach the GOE statistics for all non-vanishing dimerizations.

\subsection{Dependence on the interaction strength}

\begin{figure}[h]
	\centering
	\includegraphics[width = 0.48\textwidth]{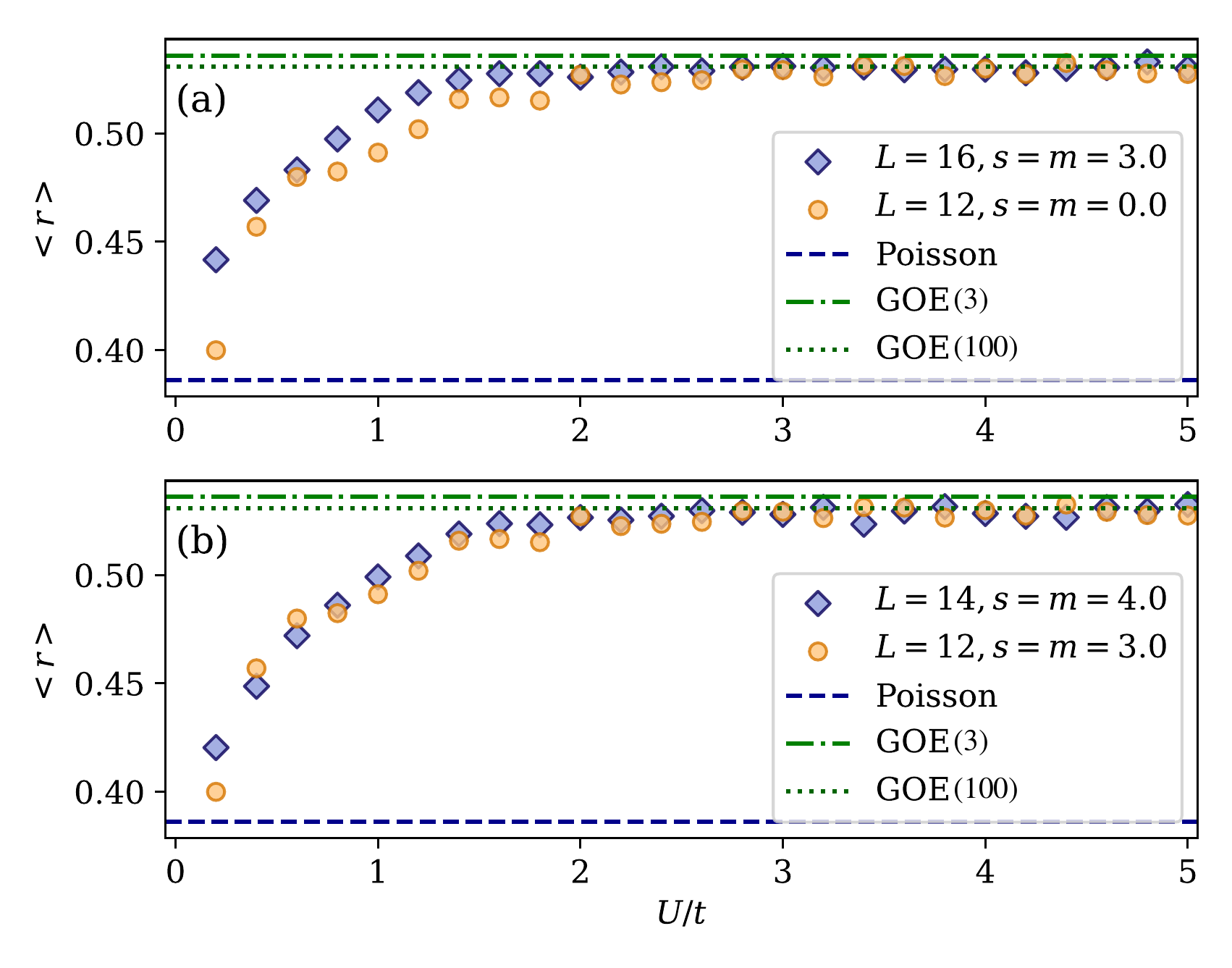}
	\caption{Mean values of the distribution of ratios $P(r)$ for $L = 12,14$ for (a) quarter and (b) half filling in dependence on the interaction $U/t$ with $t'/t = 2$ fixed. For both system sizes the $k$, $p$ sector statistics are merged and we use $\xi = 0$.}
	\label{fig:stagHubb_mr_U}
\end{figure}

As conducted for the hopping amplitude we consider the influence of the interaction strength on the mean value $\langle r\rangle$ of the distribution of ratios $P(r)$ in figure \ref{fig:stagHubb_mr_U}. We again consider quarter and half filling with the same system sizes and spins. Moreover, we choose $t'/t = 2$ for a representative value. For small interaction the mean value is close to the value expected for the Poisson distribution. The deviation from the prediction of the GOE ensemble is expected since at $U=0$ the system is non-interacting and can be solved exactly.  We observe that the mean value $\langle r\rangle$  approaches the GOE prediction with increasing interaction strength. Furthermore, when increasing the system size, typically the mean ratios $\langle r\rangle$ tend to approach faster the predicted value of the GOE ensembles. However, at the considered sizes there are minor exceptions which we attribute to statistical fluctuations.

For the entire range of considered interactions we do not yet see that the deviations become again larger for large interaction size. However, we expect that for very large interaction strength the convergence of the distribution towards the GOE distribution becomes again slower with system size, that larger deviations would be found if the interaction strength is increased much beyond the band width. 

\section{Conclusion \label{sec:concl}}
We investigated the properties of the level statistics of the dimerized Fermi Hubbard model. In order to perform the analysis of the spectrum, we first identified the important symmetries. We found a simultaneously commuting set of symmetries given by the momentum $k$, parity $p$, spin $z$-component $m$, total spin $s$, pseudospin $z$-component $\tilde{n}$ and total pseudospin $\xi$. The $\eta$-symmetry caused a tower of equally spaced eigenvalues. The analysis of the distribution of ratios of level spacings in these symmetry subsectors agreed beside finite size effects with the random matrix predictions of the GOE ensemble. Following the hypothesis that the GOE like distribution signals chaotic systems, the dimerized Hubbard model is expected to show chaotic features. However, several symmetries have to be considered separately, which will have an important influence on the dynamics of the dimerized Hubbard model, such as the $\eta$-pairing symmetry. The towers of the equally spaced eigenstates will lead to important dynamical effects. As an extension of this work also the influence of dissipation on these symmetries would be interesting. The influence of dissipation has been used in order to generate $\eta$-pairing correlations in the dissipative Hubbard model in Ref.~\cite{BernierKollath2013}.

\section*{Acknowledgements}
We thank A.~L\"auchli for sharing his deep knowledge and his codes with us which enabled us to address this project. We further thank H. Monien, J. de Marco, L. Tolle for stimulating discussions. 
We acknowledge funding from the Deutsche Forschungsgemeinschaft (DFG, German Research Foundation) under project number 277625399 - TRR 185 (B4) and project number 277146847 - CRC 1238 (C05). Further, we acknowledge funding from the Deutsche Forschungsgemeinschaft (DFG, German Research Foundation) under Germany’s Excellence Strategy – Cluster of Excellence Matter and Light for Quantum Computing (ML4Q) EXC 2004/1 – 390534769.

\renewcommand{\theequation}{A\arabic{equation}}
\renewcommand{\thefigure}{A\arabic{figure}}
\setcounter{equation}{0}
\setcounter{figure}{0}

\appendix
\section{Implementation of the fermionic spatial symmetries}
\label{app:translation}
In this appendix we explain some important points which need to be considered when implementing the translational and reflection symmetry explicitly in the exact diagonalization method. The implementation of such symmetries for spin and bosonic systems are well described for example in Ref.~\cite{Sandvik2010}. 
In fermionic systems, the anticommutation relations of the creation and annihilation operators make the implementation of the exact diagonalization more difficult. In particular, the difficulty stems from the existence of Fermi signs when representing the Fock states. In general a state in the position basis is defined by
\begin{align*}
&\ket{n_{1,\uparrow} \dots n_{L ,\uparrow}| n_{1,\downarrow}  \dots n_{L, \downarrow}}\\ &= \prod_{i \in I}c^\dagger_{i,\uparrow} \prod_{j \in J}c^\dagger_{j,\downarrow}\ket{0}\\ &= (c^\dagger_{1,\uparrow})^{n_{1,\uparrow}} \cdots(c^\dagger_{L,\uparrow})^{n_{L,\uparrow}}(c^\dagger_{1,\downarrow})^{n_{1,\downarrow}} \cdots(c^\dagger_{L,\downarrow})^{n_{L,\downarrow}}\ket{0}
\end{align*}
with $n_{j\sigma}$ the number of particles at site $j$ with spin $\sigma$ and $I, J$ sets containing the indices of sites that contain particles. In this definition an order is imposed on the operators. This order can be arbitrarily defined, but needs to be consistent within the calculations. Exchanging two fermionic operators can lead to a sign change of the prefactor of the Fock state. 
This is what can happen when applying a symmetry transformation as e.g. a translation. The 'sign' of the resulting Fock state might change with respect to the 'naive' shift of the position of the Fermions. The considerations of the sign needed in the implementation can be performed separately for each sector of the particle number and magnetization. Often, simple rules can be established.  

\paragraph{Translation symmetry}
Let us discuss the arising signs using the translation operator by two sites which acts like
\begin{equation}
T_2 c_{j,\sigma}(T_2)^\dagger = c_{j+2, \sigma}
\end{equation}
onto the creation and annihilation operators.

The simplest case is that the two-site translation does not shift a particle over the end of the chain to the beginning of the chain. The order of the operators remains preserved and no sign appears. Also if an even number of particles of a certain spin is shifted over the end of the chain to the beginning of the chain, the corresponding operators can be moved through in pairs to restore the correct order of the operators after the application of the translation. Thus, this will not generate any signs for the Fock state. The same occurs if a Fermion is shifted from the end to the beginning of the chain, but it has to be exchanged with an even number of Fermions with the same spin. This means that for the two-site translation $T_2$ no sign arises, if $N_\sigma$ is odd. In contrast, we obtain a sign change for the case that only one Fermion is shifted over the end of the chain to the beginning and $N_\sigma$ is even.

\paragraph{Reflection symmetry}

The reflection operator acts on the operators like
\begin{equation}
R c_{j, \sigma} R^\dagger = c_{L+1-j, \sigma}.
\end{equation}
Applying this on a state reverses the order of the creation operators. When commuting back we obtain a Fermi sign in the respective spin sectors if $(N_\sigma^2-N_\sigma)/2$ is odd.
Thus, in the implementation which uses the representation by Fock states, one needs to take care of these Fermi signs arising. 

\bibliographystyle{apsrev4-2}

\end{document}